\documentclass{article}
\usepackage{authblk}
\begin{document}
\title{The Habitability of Large Elliptical Galaxies}
\author{Daniel P. Whitmire}
\affil{Department of Mathematics, The University of Arkansas, Fayetteville, AR}

\date{}
\maketitle

\begin{abstract}
Based on numbers of stars, supernova rates, and metallicity, a prior study (Dayal et al. 2015) concluded that large elliptical galaxies contain up to 10,000 times more habitable planets than the Milky Way and are thus the "cradles of life". Using the results of their model and taking into account galactic number distributions and supernova rates I argue here that this result constitutes a
violation of the Principle of Mediocrity as applied to the reference class of all extant technological species. Assuming that we are a
typical technological species in the attribute of inhabiting a relatively large disk-dominated galaxy, I outline two hypotheses that
could significantly  limit the habitability of large elliptical galaxies: (1) massive  galactic sterilization events associated with quasar/AGN activity and starburst supernovae that occurred when the antecedents of today's large elliptical galaxies were much more compact; and (2) the probability of habitable planet formation in large elliptical galaxies may be small since a  disproportionately larger number of gaseous  planets are expected to form as a result of the generally higher metallicity in large elliptical galaxies. Consequently, fewer habitable planets will accrete if the gaseous planets inward migrations are sufficiently slow. The sterilization events of
Hypothesis (1) occurred at earlier epochs (z $\geq$ 1) and so they must be effectively permanent, implying two possible scenarios regarding the origin and evolution of life. In connection with one of these scenarios, independent  applications of the Principle of Mediocrity  suggest that M-dwarf stars are not significant hosts of technological life.

\end{abstract}

\section{Introduction}

In this paper I consider the physical implications of the application of the Principle of Mediocrity (POM) to the conclusion that there are significantly more habitable planets in large spheroid-dominated galaxies than in disk-dominated galaxies. This conclusion is based in part on the analysis of Dayal et al.
(2015), who found that giant elliptical galaxies have up to $10^4$ times more habitable planets than the Milky Way. In their paper they  note that the Milky Way is a typical galaxy in the
phase space of stellar mass, star formation rate, and metallicity (Fig. 1). In these attributes the Milky Way is consistent with the POM as applied to the reference class of all galaxies. However, I will argue that given Dayal et al.'s analysis, convolved with the distribution of disk-dominated and spheroid-dominated galaxies, the
Milky Way is inconsistent with the POM as applied to the reference class of all technological species.
I then consider two hypotheses which, if either is correct, will remove this inconsistency.

The application of the POM to infer physical phenomena is not new. James Gregory, Isaac Newton and Christian Huygens estimated the distance between Earth and the star Sirius by assuming that the sun is a typical star (Gingerich 2006; https://en.wikipedia.org/wiki/Mediocrity). More recently, the allies in World War II statistically estimated the number of new German Panther tanks from a small sampling of serial numbers from captured or destroyed tanks and parts (https://en.wikipedia.org/wiki/German\_tank\_problem). The assumption was that the serial numbers were sequential and random. Thus a serial number of 5 would be more typical of a total number of $\sim$ 10 tanks than say $\sim$ 1,000 tanks. After the war the relative accuracy of this POM approach was confirmed from German documents. In contrast to these two applications, the POM reference class of interest here is intelligent observers or technological species. The POM and this reference class is considered in more detail in Whitmire (2019a, 2019b). For brevity I will refer to the application of the POM using the reference class of all extant technological species as the Anthropic Principle of Mediocrity (APOM), with the caveat that technological species need not be human.

The paper outline is as follows. The Dayal et al.  results are reviewed in Section 2 and then the degree to which the APOM is invalidated is estimated. Finding that our Galaxy is  inconsistent with the APOM, two
mitigating hypotheses are considered. In Section 3 I consider Hypothesis (1): The sterilization of habitable planets by quasar/AGN activity
and starburst supernovae at earlier epochs when these spheroid galaxies were more compact. Section 4 considers the implications of the requirement that the sterilizations must be effectively permanent. In Section 5 I consider Hypothesis (2): Habitable planet formation is suppressed in large elliptical galaxies. In Section 6 the relevance of M-dwarf habitability is noted and two separate applications of the APOM are given which suggest that these most numerous
stars are unlikely hosts to technological life. Section 7 is a summary and concluding remarks.

\section{Argument that large elliptical galaxies are the "cradle of life"  and the implied improbability of being in a disk-dominated galaxy}

Dayal et al.  (2015) modelled the number of habitable terrestrial planets in local (z $\approx$ 0) galaxies relative to that of the Milky Way. They
conclude that large elliptical galaxies contain up to $10^4$ times more habitable planets than the Milky Way.  These galaxies are sometimes referred to as "red and dead"
since their star formation rates (SFR) are minimal and therefore they are under-abundant in young massive  blue stars. Consequently their current
supernova rate is very low. Dayal et al.  consider three factors as a measure of the number of habitable planets: (1) the total stellar mass, (2) the supernova rate, which
is related to the SFR  through the initial mass function, and (3) the average metallicity assuming that the stellar metallicity is the same as the
observed gas.  These three factors are related through the "fundamental metallicity relation" (Laura-Lopez et al. 2010; Maunnucci et al. 2010) which the authors exploit. This relation is based on Sloan Digital Sky Survey observations of 140,000 galaxies in the local universe. In their modelling some uncertainties such as the critical kill distance from a supernova are factored out of their final results since the number of habitable terrestrial
planets is not absolute but relative to the Milky Way.

Dayal et al.'s result for the relative number of habitable  terrestrial planets is given by
\begin{equation}
N_H \propto \frac{M^{2} Z_{g}^{\alpha}}{\psi} \; \propto \frac{M^{2}}{\psi}
\end{equation}
where $M$ is the galactic stellar mass, $\psi$ is the star formation rate, $Z_g$ is the gas metallicity, and $\alpha$ is a metallicity parameter which
is taken to be either 0 or 1 in order to study the sensitivity of their results to uncertainties in metallicity. Ultimately they assume $\alpha$ = 0
for terrestrial planets since their formation is fairly independent of metallicity over the range 0.25 to 2.5 that of the solar value (Buchhave et al.
2012; Petigura et al.  2013).  Low mass galaxies (less than $\sim 10^{9}$M$_{\odot}$) make a negligible contribution to $N_H$ for any star
formation rate (Dayal et al. 2015). (However see Stojkovi\'c et al. (2019) who propose a bimodal model for galactic habitability in which metal rich dwarfs also contribute.)

Dayal et al.  conclude that elliptical galaxies with stellar mass greater than $\approx 10^{11}$M$_{\odot}$ (about twice the mass of the Milky Way) and star formation rates $\leq$ 0.1 that of the Milky Way
have $\approx$ 10$^2 - 10^4$ times more terrestrial planets than the Milky Way. That range can be estimated from Equation (1) by taking ($M/M_{MW}) \approx$ 2 - 10 and $(\psi/\psi_{MW})\approx 0.01 - 0.10$. This  comparison is between the Milky Way and a specific given large elliptical galaxy (or a range of galaxies). To determine if there is a general conflict with the APOM we need to also consider the  number densities of elliptical and spiral
galaxies, or more generally the number densities of spheroid-dominated and disk-dominated galaxies (Kelvin et al. 2014; Moffett et al. 2016; Dayal et al. 2015).

Zackrisson et al. (2016) have considered this same question. By convolving their results for the total fraction of terrestrial planets as a function of galactic mass with the number distribution of spheroid-dominated and disk-dominated galaxies given by Kelvin et al. (2014) they conclude that "
... if our planet were randomly drawn from the cosmic distribution of habitable terrestrial planets at z = 0, we should with $\geq$ 90\% probability expect to find ourself living in an elliptical galaxy". Alternatively, they find that the probability $P$ of finding ourselves in a disk-dominated galaxy is $\leq$ 0.10. Zackrisson et al. estimate that in terms of galactic stellar mass {\em alone} the probability of finding ourselves in a disk-dominated galaxy is 1/4 (which is consistent with Kelvin et al.'s probability of 0.3). From Equation (1) above, they then divide this by the conservative star formation rate ($\propto$ supernova rate) of $\leq 0.25$ that of the Milky Way to obtain their approximate limit of $P \leq$ 0.10. Although there are minor exceptions, the typical SFR in local large elliptical galaxies is very small. If we instead divide Equation (1) by the  range of SFR's given above  ($\psi \sim 0.01 - 0.10$) then $P \sim 0.0025 - 0.025$ or roughly $\sim 0.01$, suggesting a more significant conflict with the APOM. Though somewhat subjective, Zackrisson et al. believe that even the probability limit of $P \leq 0.10$ is a problem for the Principle of Mediocrity and sufficient to warrant further study. In passing they suggest that maybe the harmful effects of supernovae were overestimated. However, Dayal et al.'s results are all relative to the Milky Way and so the absolute lethal distance from a supernova is not relevant. The improbability estimate $P \sim 0.01$ is not highly significant but sufficient to justify consideration of possible physical explanations. Though motivated by this inconsistency with the APOM, the physical explanations discussed below stand alone as necessary considerations in evaluating the habitability of large elliptical galaxies.

These results beg the question (first posed by Zackrisson et al.) - Why do we not find ourselves located in a large elliptical galaxy? The modelling of Dayal et al. appears reasonable,
especially since the result for the number of habitable planets is relative to the Milky Way.
There is one possible loophole however. Their conclusion is based on observations of the local universe. Today's large quiescent elliptical galaxies have a very different
evolution than disk-dominated galaxies like the Milky Way. Observations show that today's quiescent large elliptical galaxies were once much more compact (Ferguson et al. 2004;
Trujillo et al. 2007; Fan et al. 2008; Bezanson et al. 2009; Almaini et al. 2017)
and may have experienced quasar/AGN and starburst activity in this compact state. The evolution from the compact spheroids to today's large elliptical
galaxies may have been due to gas mass loss (Fan et al. 2008) or minor dry mergers (Bezanson et al. 2009). Both models are consistent with the
hypothesis discussed below since the stellar masses of today's large elliptical galaxies are approximately the same as their progenitors. Also any
subsequent dry mergers would involve small sterile galaxies.

\section{The sterilization of large elliptical galaxies}

\subsection{Sterilization by quasar/AGN Activity}

Balbi and Tombesi (2017) investigated the habitability of the Milky Way during the active phase of it's $3.6\times 10^{6}$ M$_{\odot}$ supermassive
black hole (SMBH), Sgr A$^*$. They considered the loss of planetary atmospheres due to extreme UV and X-ray radiation and also unprotected biological
exposure. In the latter unrealistic case they found that for accretion at the Eddington limit unprotected eukaryote and multicellular organism would receive a
lethal radiation dose at distances of 13 and 8 kpc for torus optical depths of 0 and 1, respectively. Obviously significant atmospheres on planets beyond a few
kpc's would mitigate these effects and so Earth and Earth-like planets would not be affected.  Wislocka et al. (2019)  considered the atmospheric loss  of 54
known  exoplanets when Sgr A$^*$  was active. They found that planets in the Galactic bulge might  loose several Earth atmospheres while planets beyond
7 Kpc were safe from major atmospheric erosion. They also note that Earth-like exoplanets may have lost the equivalent of 15 Mars atmospheres over
a 50 Myr  AGN period up to z = 0.5.

The SMBH's in  large elliptical galaxies like M87, $\approx 6.5\times 10^9$ M$_{\odot}$ (Akiyama et al. 2019), are expected to be  a thousand times
more massive  than Sgr A$^*$, and so the magnitude of planetary atmosphere loss and lethal radiation doses would be proportionally higher, and even at an
Earth distance of 8 Kpc significant atmospheric effects are likely. For the progenitor  compact spheroid galaxies $\approx$ 1/3 the size of M87
(Fan et al. 2008) the effects would be more severe by a factor of $\sim 10^4$ at all distances.

A more robust and general result that is directly applicable to large elliptical galaxies and their antecedents  can be obtained from Forbes and Loeb
(2018) who modelled the atmospheric loss effects due to XUV radiation from quasars and AGN activity. They give critical XUV fluences required for
significant atmospheric and oceanic mass loss. The distance $r$ from the center of the galaxy for which a given fluence will occur is given as
$$
r = 1.8 \:  \mathrm{Kpc} \left(\frac{M_{BH}}{10^{9} M_{\odot}}\right)^{\frac{1}{2}} \left(\frac{\Psi}
{10^{17}  \mathrm{erg\: cm}^{-2}}\right)^{-\frac{1}{2}}
$$
where $\Psi$ is the XUV fluence.
The nominal fluence of $10^{17}$ erg cm$^{-2}$ lies between that required to remove an Earth atmosphere ($10^{16.4}$ erg cm$^{-2})$ and an Earth ocean
$(10^{18.8}$ erg cm$^{-2})$. For a fluence of $10^{17}$ erg cm$^{-2}$ and M$_{\mathrm{BH}}$ = 6.5$\times 10^{9}$ M$_{\odot}$ (The black hole mass in
M87) $r$ = 4.6 Kpc, which is greater than the expected effective radius of the compact antecedent spheroid, {\it {\it i.e.}} less than 1/3 the local large elliptical galaxies'  effective radius (Fan et al. 2008; Almaini et al. 2017). In the case  of M87 the effective radius of the antecedent spheroid is $\approx$ 7.4 Kpc $\times$ 1/3 $\approx$ 2.5 Kpc. Further, stars in elliptical galaxies are in mostly radial orbits, unlike stars in spiral galaxies, so they will spend some time closer to the
central black hole than might be indicated by the observed effective radius.

\subsection{Sterilization by starburst supernovae}

The SFR in starburst galaxies appears to have a maximum of 30,000 times that of the Milky Way (Rowan-Robinson et al. 2018). I assume an
average SFR = 1/10 $\times$ 30,000 = 3,000 M$_{\odot}$yr$^{-1}$ during the starburst phase of the compact spheroid.  The Milky Way's SFR today is $\approx$ 3 M$_{\odot}$ yr$^{-1}$. Therefore the large elliptical galaxy progenitor would have a SFR = 3,000/3 = 1,000 times that of the Milky Way
today. I assume that the supernova rate in large elliptical galaxy progenitors is the same fraction of the SFR as in the Milky Way ($\sim$ 1/100).
For Earth the average rate R of SN occurring within distance $D$ is given by (Melott \& Thomas, 2011; Fields, 2004)
$$
R(\leq D) = \frac{2 \: \mathrm{events}}{\mathrm{Myr}} \left(\frac{D}{100\:\mathrm{pc}}\right)^3 = \frac{2\times
10^{-3}\;\mathrm{events}}{\mathrm{Gyr}}\left(\frac{D}{\mathrm{pc}}\right)^3  .
$$
Scaling this to a large elliptical galaxy progenitor with effective radius = 2.5 Kpc and neglecting morphology,
and assuming equal galactic stellar masses, yields a SN event rate of

$$
\frac{2\times 10^{-3}\: \mathrm{events}}{\mathrm{Gyr}}\left(\frac{D}{\:\mathrm{pc}}\right)^3 \times 1,000 \times \left(\frac{8 \:\mathrm{kpc}}{2.5
\:\mathrm{kpc}}
   \right)^{3} =  \frac{65 \:\mathrm{events}}{\mathrm{Gyr}}\left(\frac{D}{\mathrm {pc}}\right)^3
$$
where 8 Kpc is the distance of Earth from the center of the Milky Way and (8 Kpc/2.5 Kpc)$^3$ is the increase in stellar density in the large elliptical galaxy progenitor relative to the Milky Way. Setting this = 1 event/0.1 Gyr (the assumed lifetime of the starburst phase) gives the closest SN  distance
from a random star in a large elliptical galaxy progenitor of $D$ = 0.53 pc.

By comparison, for Earth, significant atmospheric and biological damage is
expected for supernovae occurring  within 8 - 10 pc (Ellis \& Schramm 1995;
Gehrels et al. 2003; Fields, et al. 2008) or possibility further (Melott et al. 2018). The rough estimate made here is only to show plausibility,
though I note that some of the assumptions
are conservative, such as the assumption that the stellar mass of the Milky Way and large elliptical galaxy progenitor are equal. In fact local large elliptical galaxy stellar masses are typically greater than the Milky Way stellar mass.  Observations indicate that the stellar masses of the
progenitors are similar (Almaini et al. 2017; Fan et al. 2008).  On the other hand, the Milky Way SN rate is the result of a disk geometry
and, all other factors the same, the rate within a given distance $D$ would be less for the same stellar mass distributed isotropically. The nearest SN
distance is fairly insensitive to uncertainties in the approximations.  If the average lifetime of the starburst phase = $10^7$ yr rather than $10^8$
yr, the closest SN to a random star would be 0.53 pc $\times 10^{1/3}$ = 1.1 pc. If the average SFR in the compact large elliptical galaxy progenitor was
1\% of the maximum = 300 rather than 3,000 M$_{\odot}$yr$^{-1}$ the closest SN would also be at a distance of 1.1 pc. If both the starburst phase
lifetime = $10^7$ yr and the SFR = 300 M$_{\odot}$yr$^{-1}$ the nearest SN during the starburst phase would still be 2.5 pc.

I note that if, contrary to current models cited here, some large elliptical galaxies form by major mergers of comparable mass galaxies with their
associated SMBH's, sterilization might still occur if the mergers happen early (as expected) when the galaxies still contained a large amount of gas.
However, this scenario will not be considered further here.

\section{Implications of the requirement that the sterilization be effectively permanent}

\subsection{Scenario 1}
Quasar/AGN activity and/or starburst supernovae must not only sterilize any existing life but also leave the planets effectively uninhabitable.
As discussed above, planetary atmospheres may be totally destroyed in the quasar/AGN phase of large elliptical galaxy's compact progenitors. Geologic
out-gassing might eventually produce another atmosphere but, if the origin of life is sensitive
to conditions in the primordial atmosphere and/or oceans, those conditions might never repeat.

\subsection{Scenario 2}
The anthropically-unfiltered timescale for the origin of prokaryote life is unknown. The fact that life evolved relatively soon on Earth ($\approx$ 3.8 Gyr ago) is
only weak evidence that the unfiltered timescale is short since our existence places a strong anthropic selection effect on this observed time (Spiegel
\& Turner, 2012). If the unfiltered typical timescale for the origin of life was $\gg$ 0.7 Gyr it would still have to occur in less than $\approx$ 1
Gyr in order that our technological species evolve before the end of the animal habitable biosphere in $\approx$ 1.3 Gyr (Wolf \& Toon 2015). Further,
the unfiltered timescale for eukaryote life to evolve from prokaryote life might be $\gg$ 1.8 Gyr, as observed on Earth. It is known that this event
occurred only once in 4 Gyr (Blackstone, 2013).  Carter and McCrea (1983) have effectively argued that there is one or possibility two
rare events in the origin of life for which the (unfiltered) timescale is much greater than the lifetime of the sun. Assuming that the origin of life
could still happen post-sterilization, a delay of several Gyr would preclude the evolution of a technological species (for solar type stars, but see
Section 6 below) assuming the unfiltered observed timescales for prokaryote and eukaryote life are the same as (or more likely) longer than observed for
Earth. I focus on these two timescales since other timescales like the evolution from eukaryote life to multicellular life occurred in a shorter amount
of time (1.1 Gyr) but more importantly it occurred independently several times (Grosberg \& Strathmann 2007; Parfrey \& Lahr 2013), suggesting its
(convergent) evolution unfiltered timescale is comparable to that observed and not much greater as is possible/likely for prokaryote and eukaryote
timescales.

\section{Suppression of habitable planet formation due to a disproportionate number of gaseous planets}
According to Dayal et al. (2015) the relative number of gaseous planets in large elliptical galaxies  is up to $10^6$ times that of the Milky Way. Thus
there is a disproportionate number (by a factor of 100) of gas planets relative to terrestrial planets. The greater proportion of gas planets in large elliptical galaxies relative to the Milky Way is due to the generally higher metallicity in large elliptical galaxies
and the fact that unlike terrestrial planets the probability of formation of gas planets depends more strongly on metallicity, $\propto 10^{2z_{gas}}$ (Dayal et al. 2015; Mortier et al. 2013). The  gas planets likely formed beyond the water condensation radius ($\approx$ 5 AU for solar type stars) and migrated inward. If the
migration is relatively slow (greater than the disk lifetime of $\sim 10^6$ yr) then these planets will force planetesimals  in
habitable zones inward, preventing the formation of future habitable planets (Izidoro, Morbidelli \& Raymond 2014). This would also apply to the Milky Way as well, though at a reduced frequency. It leads to the expectation of the
absence of Earth-like habitable planets in systems with interior gas planets.  Earth-mass planets in the
habitable zones of these systems are likely to be extremely volatile rich and therefore would not be Earth-like (Izidoro et al. 2014).

\section{APOM evidence that M-Dwarf stars are not good candidates for technological life}

 Scenario (2) above applies only to solar-type stars. Whitmire \& Matese (2009) showed that although our sun is more massive  and luminous than 95\% of
 main sequence stars, it is likely typical of stars that harbour a planet with technological life, {\it i.e.} it is consistent with the APOM.
 Terrestrial planet accretion simulations  (Raymond, Scalo \& Meadows 2007) and simple heuristic models show that the probability of accretion of a planet of mass $\geq 0.3 M_{\oplus}$  in the HZ is strongly dependent on the stellar mass, $M$, typically $\propto M^{3}$, which disfavors low mass M-stars. The probability is assumed to be proportional to the amount of solid material in the HZ which scales with disk mass which in turn scales with stellar mass. Convolving this with the mass distribution of main sequence stars (which disfavors massive stars) results in a sharp peak at $\approx$ 1 $M_{\odot}$.  Imposing the self selection condition that the stellar lifetime must be greater than the time of appearance of a technological species (which further disfavors higher mass stars) modifies  the distribution at higher stellar masses. Ultimately, we found that for a technological species that requires 4.5 Gyr to evolve the optimum stellar type is G5 with 2/3 of stars most likely to host  technological life lying between G0 - K5 stellar types. Assuming other plausible timescales  for the evolution of technological life only moderately changes this result (Whitmire \& Matese 2009).

It might be observed that M-dwarf stars have extremely long lifetimes and so in the future they could
dominate as sites of technological life. However, if true this would violate the APOM in a manner analogous to that of large elliptical galaxies, {\it
i.e.} why do we not find ourselves around an old M-dwarf star trillions of years from now? This question has been considered in detail by Loeb et al.
(2016), where the relative likelihood of life as a function of cosmic time was considered. Their conclusion is that unless life is suppressed on planets
around M-dwarfs our existence is highly premature. They argue conservatively that the probability of our existence at or before the present time is at
most 0.1\% and probably less, and comment that "sometimes rare events happen". A similar conclusion can be inferred from Dayal {\it et al.} (2018). In
the context of the APOM this can be interpreted as strong statistical evidence that technological life is rare around M-dwarfs, since otherwise our existence at the present time would be highly improbable, $\leq 10^{-3}$. According to Loeb et al. (2016) the habitable planet formation rate around solar type stars has a peak near the present epoch (A similar conclusion can be found in Livio 1999, where it is assumed that habitability traces carbon production). Thus we exist at a typical time in the evolution of the habitable universe, but only if M-dwarf stars are not significant hosts to technological life.

This APOM statistical inference is backed up by already known physical mechanisms suggesting that technological life, or even complex life,
may be difficult on tidally locked  rocky planets in the habitable zones of M-dwarf stars. For example, strong UV emissions and flares  could strip away
or erode the atmospheres of rocky planets in the habitable zones (Tilley et al. 2019). An atmosphere $\geq$ 1/10 that of Earth is necessary to adequately transfer heat from
the sun-facing hemisphere to the night hemisphere (Joshi et al. 1997). Tidal heating of planets in the habitable zones of M-dwarfs may lead to "tidal
Venuses" (Barnes et al. 2013), thus precluding technological life.


\section{Conclusion}

Assuming the analysis of a previous study and the number density distributions of disk-dominated and spheroid-dominated galaxies, the probability that a randomly selected technological species would be in a disk-dominated galaxy is estimated to be $\sim$ 0.01. This implies that we are atypical in residing in a disk-dominated galaxy, contrary to the expectation of the Principle of Mediocrity as applied to the reference class of all technological species (APOM).

Two hypotheses which could mitigate this inconsistency were considered: (1) Habitable planets in large elliptical galaxies have been previously sterilized by XUV radiation during the galaxy's quasar/AGN and starburst phases when the antecedents of today's large spheroid-dominated galaxies were more compact than their present-day passive descendants;
and (2) habitable planet formation is suppressed in large spheroid-dominated galaxies due to the expected larger numbers of  gaseous planets, whose slow inward migration could inhibit terrestrial planet accretion in stellar habitable zones.

The sterilizations of Hypothesis (1) must be
effectively permanent, which implies one of two scenarios. Scenario (1): The conditions for the origin of life exist only during early epochs in the evolution of
potentially habitable planets. Scenario (2): A delay of several Gyr doesn't allow sufficient time for a technological species to
evolve in the lifetime of the planet's animal biosphere, for solar type stars.
In connection with Scenario (2), which applies only to solar type stars, two other applications of the APOM were given
which  predict that  M-dwarf stars are not significant hosts of technological
species, a conclusion consistent with already known physical constraints on habitability for these stars. If true, this implies that we exist at a typical time in the history of the technologically habitable universe, as expected.

Lastly I note that although motivated by an inconsistency with the APOM, the physical implications discussed here stand alone as necessary considerations in evaluating the habitability of large elliptical galaxies.



{}

\end{document}